# Electrical and Thermal transport studies of Sr and Mn co-substituted NdCoO$_3$

Ashutosh Kumar*

Department of Physics, Indian Institute of Technology Patna, Bihta-801106, Bihar India

**Abstract:** Oxide thermoelectrics are exciting due to their chemical and thermal stability at high temperatures. However, the efficacy of these materials are limited by poor figure of merit (zT). In this study, the role of Sr and Mn co-substitution on the thermoelectric properties of NdCoO$_3$ (Nd$_{1-x}$Sr$_x$Co$_{1-y}$Mn$_y$O$_3$; $0.00 \leq x \leq 0.10$; $0.00 \leq y \leq 0.10$) is investigated. The Seebeck coefficient decreases with single Sr substitution at Nd site; however, the Sr and Mn co-substitution enhances the Seebeck coefficient compared to single Sr substitution and is attributed to the localization effect. Sr substitution at La site creates hole in the system and results in enhanced electrical conductivity ($\sigma$); however, $\sigma$ reduces with Mn substitution at Co site in NdCoO$_3$. A reduced thermal conductivity for the co-substituted samples is observed and attributed to decrease in phonon thermal conductivity. Simultaneous optimization of TE parameters results in improved zT ~ 0.038 for Nd$_{0.95}$Sr$_{0.05}$Co$_{0.95}$Mn$_{0.05}$O$_3$ at 540 K.

*Email: science.ashutosh@gmail.com
Present Address: Lukasiewicz Research Network-Krakow Institute of Technology, Zakopianska-73, Krakow 30-418, Poland

## Introduction

Oxide materials are exciting candidate for power generation applications due to their chemical and thermal stability at higher temperatures and in open environments.[1,2] However, they are limited by poor energy conversion efficiency due to insufficient figure of merit (zT). In general, the efficacy of a TE material depends on its physical parameters: Seebeck coefficient ($\alpha$), electrical conductivity ($\sigma$), thermal conductivity ($\kappa$) as $zT=\alpha^2\sigma T/\kappa$, where $\alpha^2\sigma$ is known as power factor and $T$ is absolute temperature. A promising TE materials requires high $\alpha$ and $\sigma$ values along with low $\kappa$; however, it is quite challenging to attain these requirement in a single system due to their interdependence. The rise in $\sigma$ leads to reduced $\alpha$ due to their opposite dependence on carrier concentration and hence an optimized $\alpha^2\sigma$ is required. Also, total thermal conductivity ($\kappa$) is a combination of two components: electronic thermal conductivity ($\kappa_e$) and phonon thermal conductivity ($\kappa_{ph}$). Increase in $\sigma$ also results in simultaneous rise in $\kappa_e$ according to Wiedemann Franz law. Hence, the simultaneous optimization of $\alpha^2\sigma$ and $\kappa_{ph}$ may results to an improvement in zT. There are several methodology employed to enhance zT value in oxide TE systems either by improving $\alpha^2\sigma$ and/or reducing $\kappa_{ph}$ [3–7]. The increase in $\sigma$ in oxide systems are realized by proper substitution at cation sites as well as using composite approach that includes conducting second phase that enhances the electron/hole concentration in the system. Further, promising strategies like nanostructuring, lattice defects, mass fluctuations, composite approach etc. enhances the phonon scattering benefical for lowering of $\kappa$.[5,8–15] Kumar et al. recently developed experimentally the simultaneous optimization of electronic and thermal transport properties in oxide and chalcogenide materials for improved zT. [16–18].

Cobalt-based oxide materials (e.g. $LaCoO_3$, $NaCoO_2$, $Ca_3Co_4O_9$) are capable to become one of the best TE system due to the presence of different charge states ($Co^{2+}$, $Co^{3+}$, $Co^{4+}$) as well as corresponding different spin states $Co^{3+}$:LS ($t_{2g}^6 e_g^0$), IS ($t_{2g}^5 e_g^1$), HS ($t_{2g}^4 e_g^2$) and $Co^{4+}$:LS ($t_{2g}^5 e_g^0$), IS ($t_{2g}^4 e_g^1$), HS ($t_{2g}^3 e_g^2$) where LS, IS and HS are low spin, intermediate spin and high spin respectively.[19,20] The presence of these charge states along with the spin-states affect $\alpha$ and electrical conductivity following Heikes formula.[21] Also, these spin-states are temperature-dependent and also the transition temperature in the $LaCoO_3$ based compounds can be tuned with the alkaline metals substitution at La site.[22] There are several studies reported in literature for La-based cobaltates [23–30]; however, the other rare-earth elements are not well investigated. Since the transport properties in oxides are quite exciting with different rare-earth elements. Hence, it is important to explore the TE properties in other rare-earth cobaltates. Recently, Dudnikov et al. showed the TE properties in parent $NdCoO_3$ and $SmCoO_3$ systems [31]; however, the improvement in TE properties are not obtained. In this report, we show the TE properties in Sr-and Mn co-substituted $NdCoO_3$ system over a wide temperature range from 300 K-773 K.

## Experimental Section

Polycrystalline samples of $Nd_{1-x}Sr_xCo_{1-y}Mn_yO_3$, $0.00 \leq x \leq 0.10$; $0.00 \leq y \leq 0.10$ were synthesized using standard solid-state method. In a typical synthesis process, stoichiometric amount of $Nd_2O_3$, $SrCO_3$, $Co_3O_4$ and $Mn_2O_3$ (All precursors are of Sigma Aldrich with purity greater than 99.9 %) were taken and mixed thoroughly using mortar-pestle. The mixtures were calcined in a muffle furnace using the alumina crucibles at 1473 K for 24 hours with 3°/min heating and cooling rate. Further, the calcined powders were again ground thoroughly in mortar-pestle for 2 hours and then consolidated into cylindrical shape having 10 mm in diameter and 12 mm in height and were sintered at 1523 K for 12 hours in a muffle furnace. The sintered pellets were cut to proper dimensions using a precise wire saw for further measurements. Crystallographic structure and phase identification were done *via* x-ray diffraction (XRD) pattern using Rigaku diffractometer ($\lambda$=1.5406 Å) followed by Rietveld refinement [32]. Field emission scanning electron

*Email: science.ashutosh@gmail.com
Present Address: Lukasiewicz Research Network-Krakow Institute of Technology, Zakopianska-73, Krakow 30-418, Poland

microscopy (FE-SEM, Zeiss) was performed for microstructural investigations. Electrical conductivity and Seebeck coefficient are measured using the SBA 458 (NETSCH) apparatus in four probe configuration in a wide temperature range from 300 K to 773 K. Thermal conductivity (κ) measurement was performed using the following equation: $\kappa = D\rho C_p$. The thermal diffusivity (D) is measured using laser flash technique, specific heat capacity ($C_p$) is calculated using Dulong-Petit law, and bulk density (ρ) is calculated using the sample mass and its geometric volume. The scanning thermoelectric microscope (STM) measurement was used to probe the spatial distribution of Seebeck coefficient for all the samples at 300 K.

## Results and Discussion

Figure 1 depicts the powder x-ray diffraction (XRD) pattern for $Nd_{1-x}Sr_xCo_{1-y}Mn_yO_3$, 0.00≤x≤0.10; 0.00≤y≤0.10. All the diffraction pattern shows similar nature and indicates a single phase formation. No trace of impurity is observed within the sensitivity of the XRD. The diffraction pattern consists of characteristic 2θ peak at 33.60°. Further, the diffraction pattern obtained for $NdCoO_3$ is indexed in accordance to a orthorhombic phase with *Pbnm* space group and is found to be in agreement with the ICDD file no. 074-8952.

The powder diffraction pattern is further analyzed using Rietveld refinement method employing Fullprof™ software. The refinement of the $Nd_{1-x}Sr_xCo_{1-y}Mn_yO_3$ samples has been done in orthorhombic crystal structure and *Pbnm* space group. The following Wyckoff positions are used for the refinement: For Nd, 4c (x, y, 1/2) for Co, 4b (1/2, 0, 0) and there are two oxygen cites 4c (x, y, 1/2) and 8d (x, y, z). The lattice parameters of all the samples along with refinement parameters are shown in Table I. The refinement parameters are well in the acceptable limit. It is observed that the addition of Sr at La site leads to increase in the lattice spacing and is attributed to the larger ionic radii of Sr in comparison to La. Further, Mn substitution at Co site also leads to increase in the lattice parameters.

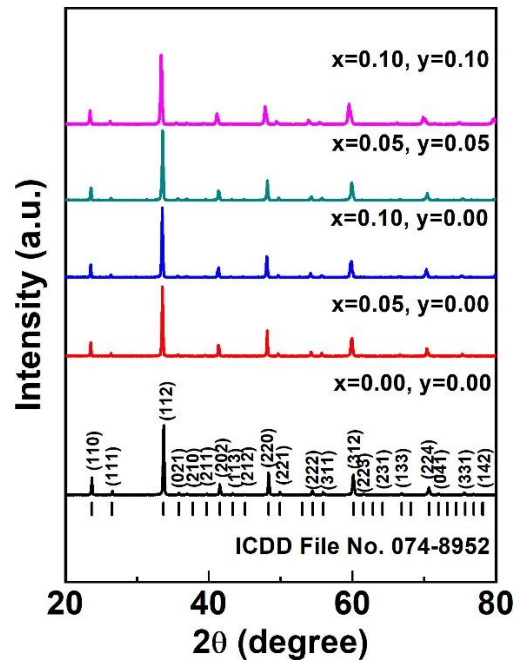

Figure 1. Powder X-ray diffraction pattern for $Nd_{1-x}Sr_xCo_{1-y}Mn_yO_3$ polycrystalline samples. Miller indices and Bragg's position for $NdCoO_3$ is marked.

*Email: science.ashutosh@gmail.com
Present Address: Lukasiewicz Research Network-Krakow Institute of Technology, Zakopianska-73, Krakow 30-418, Poland

Table I: Refinement parameters $R_{wp}$, $R_{exp}$, $R_p$, $\chi^2$, lattice parameters (*a, b, c*) and volume of the unit cell for $Nd_{1-x}Sr_xCo_{1-y}Mn_yO_3$, $0.00 \leq x \leq 0.10$; $0.00 \leq y \leq 0.10$ sample calculated from the Rietveld refinement of the XRD patterns.

| Sample | *a* (Å) | *b* (Å) | *c* (Å) | $R_{wp}$ | $R_{exp}$ | $R_p$ | $\chi^2$ |
|---|---|---|---|---|---|---|---|
| x=0.00, y=0.00 | 5.3443 | 5.3298 | 7.5498 | 19.1 | 15.1 | 11.2 | 1.60 |
| x=0.05, y=0.00 | 5.3568 | 5.3368 | 7.5598 | 18.5 | 14.3 | 11.4 | 1.68 |
| x=0.10, y=0.00 | 5.3767 | 5.3444 | 7.5716 | 19.5 | 14.5 | 11.5 | 1.81 |
| x=0.05, y=0.05 | 5.3624 | 5.3400 | 7.5569 | 18.8 | 15.6 | 10.2 | 1.45 |
| x=0.10, y=0.10 | 5.3854 | 5.3717 | 7.6061 | 17.3 | 13.6 | 10.3 | 1.62 |

Field emission scanning electron microscope images of $Nd_{1-x}Sr_xCo_{1-y}Mn_yO_3$, $0.00 \leq x \leq 0.10$; $0.00 \leq y \leq 0.10$ samples are shown in Fig. 2(a-e). The crystallite size for these samples lies in the range of 1-2 μm and is attributed to the higher calcination temperature chosen during synthesis. Sr and Mn substitution in $NdCoO_3$ shows similar surface morphology.

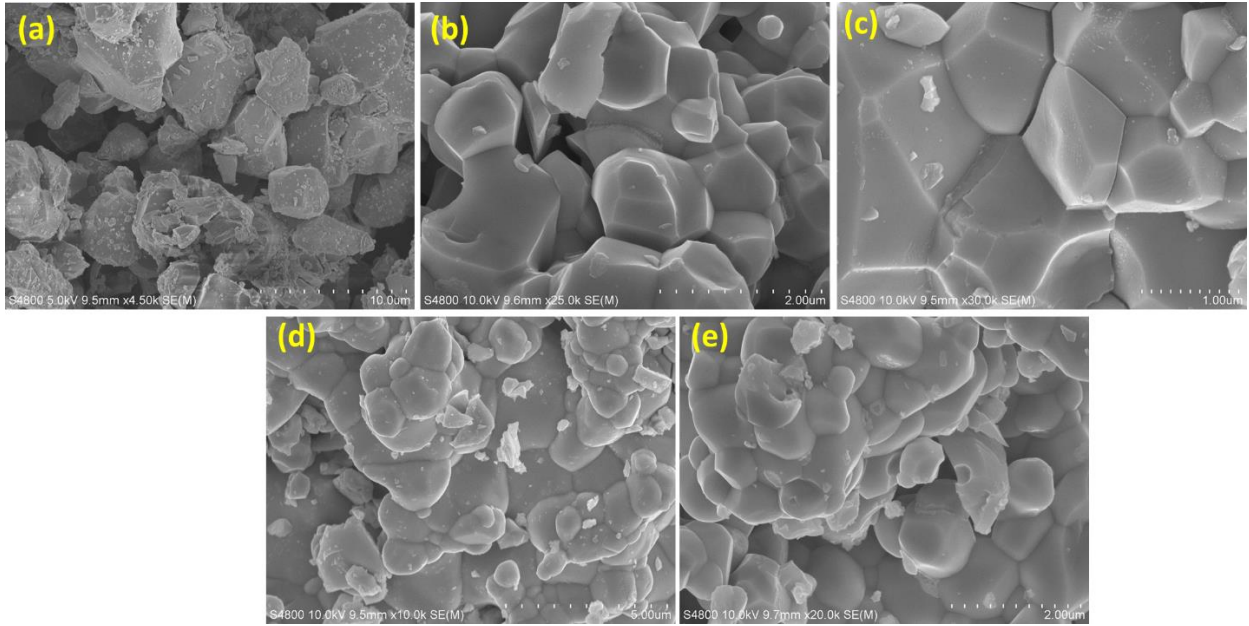

Figure 2. FESEM image for $Nd_{1-x}Sr_xCo_{1-y}Mn_yO_3$ (a) x=0.00, y=0.00 (b) x=0.05, y=0.00 (c) x=0.10, y=0.00 (d) x=0.05, y=0.05 (e) x=0.05, y=0.10 (f) x=0.10, y=0.10.

Figure 3(a) depicts the temperature-dependent Seebeck coefficient (α) for $Nd_{1-x}Sr_xCo_{1-y}Mn_yO_3$, $0.00 \leq x \leq 0.10$; $0.00 \leq y \leq 0.10$ over a wide temperature range from 300-773 K. The pristine $NdCoO_3$ shows negative α value near room temperature indicating a dominating n-type conduction in the system, which may indicate the presence of electrons. Similar values of Seebeck coefficient is obtained in earlier reports in $NdCoO_3$ system.[31] The value of α for the sample with x=0.00, y=0.00 is -200 μV/K at 300 K. The alpha changes sign with an increase in temperature and shows a maximum value at 448 K (~250 μV/K) and then decreases with further increase in temperature. The α decreases with an increase in temperature for all the samples. Further, the Sr and Mn subsituted samples shown positive values of α indicating the dominating nature of holes in the substituted samples. This change in sign may be attributed to the creation of hole with Sr substitution at La site. The Seebeck coefficient (α) decreases from 225 µV/K for x=0.05, y=0.00 to 170 µV/K for x=0.10, y=0.00 at 300 K. However, Sr and Mn co-substitution at Nd and Co site show improved α compared to only Sr substitution [240 µV/K for x=0.05, y=0.05 at 300 K].

*Email: science.ashutosh@gmail.com
Present Address: Lukasiewicz Research Network-Krakow Institute of Technology, Zakopianska-73, Krakow 30-418, Poland

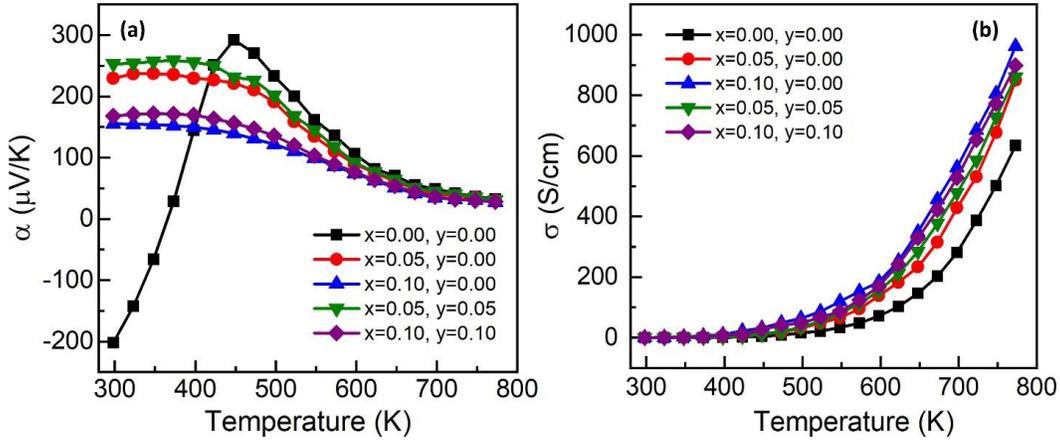

Figure 3. Temperature-dependent (a) Seebeck coefficient (α), (b) Electrical conductivity (σ) as a function of temperature for $Nd_{1-x}Sr_xCo_{1-y}Mn_yO_3$, $0.00 \leq x \leq 0.10$; $0.00 \leq y \leq 0.10$.

Figure 3(b) shows the temperature-dependent electrical conductivity (σ) of $Nd_{1-x}Sr_xCo_{1-y}Mn_yO_3$, $0.00 \leq x \leq 0.10$; $0.00 \leq y \leq 0.10$ samples. The σ of all the samples increases with the increase in temperature showing the semiconducting behavior. Also, the Sr substitution at Nd site leads to improve σ. The increase in σ may be attributed to the creation of hole due to Sr substitution at La site. The Mn substitution at Co site may results in the localization of charge carriers due to possible bonding between their different charge states.[23,33] The increase in σ with the increase in Sr and Mn substitution in $NdCoO_3$ is consistent with the decrease in α. The σ at 300 K for $NdCoO_3$ is 0.002 S/cm and it further increases with the Sr substitution at Nd site. The σ at 773 K increases from 650 S/cm for x=0.00, y=0.00 to 1000 S/cm for x=0.10, y=0.00.

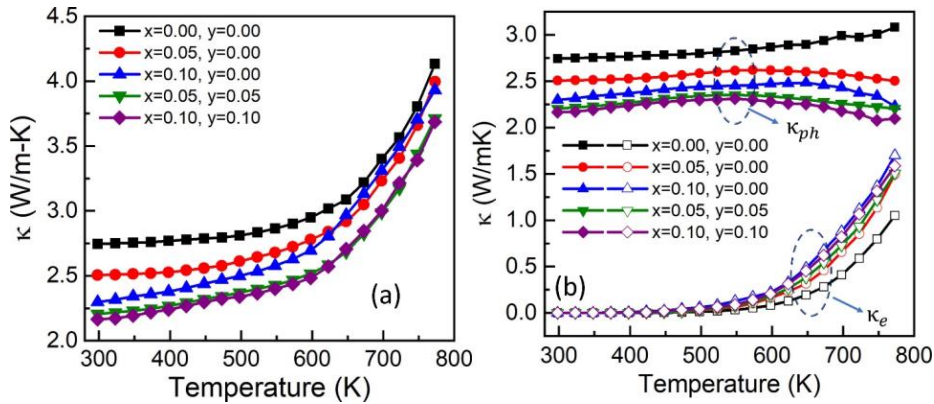

Figure 4. (a) Total thermal conductivity (κ), (b) electronic thermal conductivity ($κ_e$), and phonon thermal conductivity ($κ_{ph}$) as a function of temperature for $Nd_{1-x}Sr_xCo_{1-y}Mn_yO_3$, $0.00 \leq x \leq 0.10$; $0.00 \leq y \leq 0.10$.

*Email: science.ashutosh@gmail.com
Present Address: Lukasiewicz Research Network-Krakow Institute of Technology, Zakopianska-73, Krakow 30-418, Poland

Figure 4(a) depicts the changes in total thermal conductivity (κ) as a function of temperature of $Nd_{1-x}Sr_xCo_{1-y}Mn_yO_3$, $0.00 \leq x \leq 0.10$; $0.00 \leq y \leq 0.10$. The κ increases with increase in temperature for all the samples. However, the total thermal conductivity decreases with Sr and Mn substitution in $NdCoO_3$. The κ of $NdCoO_3$ at 300 K is 2.75 W/m$^{-1}$K$^{-1}$. It is known that κ consists of $κ_e$ and $κ_{ph}$. It is important to understand the individual nature of $κ_e$ and $κ_{ph}$ for $Nd_{1-x}Sr_xCo_{1-y}Mn_yO_3$, $0.00 \leq x \leq 0.10$; $0.00 \leq y \leq 0.10$. For the estimation of $κ_e$, Wiedemann-Franz law has been used: $κ_e = LσT$, here $L$ is the Lorenz number, $T$ is absolute temperature. The $L$ is calculated using the equation proposed by Snyder ($L=1.5+\exp[-|α|/116]$) [34]. Further, $κ_e$ is subtracted from κ to estimate $κ_{ph}$. Figure 4(b) shows $κ_e$ and $κ_{ph}$ as a function of temperature for $Nd_{1-x}Sr_xCo_{1-y}Mn_yO_3$, $0.00 \leq x \leq 0.10$; $0.00 \leq y \leq 0.10$. It is found that $κ_e$ increases with the increase in temperature for all the sample and is attributed to the rise in electrical conductivity for all the samples with the increase in temperature. On the other hand, $κ_{ph}$ increases slightly with increase in temperature for $NdCoO_3$. However, $κ_{ph}$ decreases at higher temperatures for Sr and Mn substituted samples. This may be attributed to the decrease in the mean free path at the higher temperature which reduces the corresponding phonon thermal conductivity. Further, $κ_{ph}$ also decreases with Sr and Mn substitution in $NdCoO_3$ aross the temperature range studied. This suggest that the substitution of Sr and Mn in $NdCoO_3$ enhance phonon scattering. This may be due to the mass and strain fluctuation in the sample.[35,36]

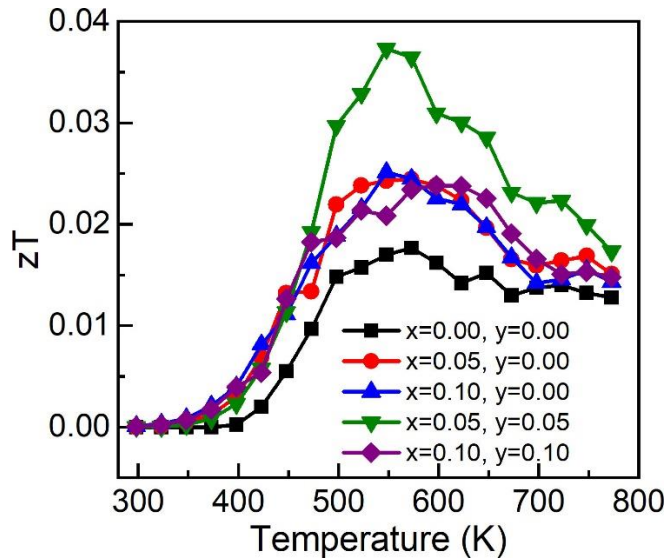

Figure 5. Figure of merit (zT) as a function of temperature for $Nd_{1-x}Sr_xCo_{1-y}Mn_yO_3$, $0.00 \leq x \leq 0.10$; $0.00 \leq y \leq 0.10$.

Using the measured values of TE parameters viz. α, σ, and κ, the TE figure of merit (zT) is calculated. The temperature-dependent zT for all the samples are shown in Fig.~5. The figure of merit for all the samples are found to increase with the increase in temperature upto 550 K and then decreases with further increase in temperature. The increase in zT is corroborated to the increase in the electrical conductivity in the system. A maximum figure of merit of 0.038 is obtained for the sample with x=0.05, y=0.05 at 540 K sample which is higher than several studies in rare earth cobalates at this temperature.[37]

*Email: science.ashutosh@gmail.com
Present Address: Lukasiewicz Research Network-Krakow Institute of Technology, Zakopianska-73, Krakow 30-418, Poland

## Conclusion

Thermoelectric properties of $Nd_{1-x}Sr_xCo_{1-y}Mn_yO_3$, $0.00 \leq x \leq 0.10$; $0.00 \leq y \leq 0.10$ samples synthesized through standard solid-state route is investigated. The structural analysis (x-ray diffraction measurement and Rietveld refinement) shows the pure phase formation of these samples, which is further supported by electron microscopy studies. Seebeck coefficient ($\alpha$) shows negative value at 300 K depicting the n-type dominating conduction in the system. However, with Sr (hole) doping at La site, the p-type conduction is observed across the wide temperature range (300-800 K). Further hole doping increases the electrical conducting at the cost of $\alpha$. The Mn substitution at Co site leads to improved $\alpha$ with moderate $\sigma$ and hence results to an improved power factor for Sr and Mn co-doped $NdCoO_3$. Thermal conductivity increases with increase in temperature and is attributed to rise in $\sigma$ with temperature. The phonon thermal conductivity decreases with temperature and is attributed to the phonon scattering at higher temperature. The optimized value of $\sigma$, $\alpha$, and $\kappa$ results in a maximum zT of 0.038 at 560 K for $Nd_{0.95}Sr_{0.05}Co_{0.95}Mn_{0.05}O_3$.

## Acknowledgement

A. Kumar thank ministry of human resource and development (MHRD) India for financial support.

*Email: science.ashutosh@gmail.com
Present Address: Lukasiewicz Research Network-Krakow Institute of Technology, Zakopianska-73, Krakow 30-418, Poland

*Email: science.ashutosh@gmail.com
Present Address: Lukasiewicz Research Network-Krakow Institute of Technology, Zakopianska-73, Krakow 30-418, Poland

*Email: science.ashutosh@gmail.com

Present Address: Lukasiewicz Research Network-Krakow Institute of Technology, Zakopianska-73, Krakow 30-418, Poland